\documentclass[letterpaper, 10 pt, conference]{ieeeconf}  

\IEEEoverridecommandlockouts                              
\usepackage[T1]{fontenc}
\usepackage{verbatim}
\usepackage{subfigure}

\usepackage[utf8]{inputenc}
\usepackage{times}
\usepackage{booktabs}
\usepackage{units}
\usepackage{url}
\usepackage{amsmath}
\usepackage{amssymb}
\usepackage{mathdots}
\usepackage{graphicx}
\usepackage{esint}
\usepackage{color}
\usepackage{wrapfig}
\usepackage[margin=1in]{geometry}
\usepackage{cancel}
\usepackage{multirow}
\usepackage{tikz}
\usetikzlibrary{decorations.pathreplacing}
\usetikzlibrary{fadings}
\usepackage{graphicx}

\title{\LARGE \bf
A Physics-Based Data-Driven Approach for Finite Time Estimation of Pandemic Growth
}

\author{Harshvardhan Uppaluru, Hamid Emadi, and  Hossein Rastgoftar
\thanks{The authors are with the Aerospace and Mechanical Engineering Department at University of Arizona. Emails: \{huppaluru, hamidemadi, hrastgoftar\}@email.arizona.edu}}

\begin{document}
\maketitle
\thispagestyle{empty}
\pagestyle{empty}

\begin{abstract}
COVID-19 is a global health crisis that has had unprecedented, widespread impact on households across the United States and has been declared a global pandemic on March 11, 2020 by World Health Organization (WHO) \cite{WHO}. According to Centers for Disease Control and Prevention (CDC) \cite{CDC}, the spread of COVID-19 occurs through person-to-person transmission i.e. close contact with infected people through contaminated surfaces and respiratory fluids carrying infectious virus. This paper presents a data-driven physics-based approach to analyze and predict the rapid growth and spread dynamics of the pandemic. Temporal and Spatial conservation laws are used to model the evolution of the COVID-19 pandemic. We integrate quadratic programming and neural networks to learn the parameters and estimate the pandemic growth. The proposed prediction model is validated through finite-time estimation of the pandemic growth using the total number of cases, deaths and recoveries in the United States recorded from March 12, 2020 until October 1, 2021 \cite{OWM}.
\end{abstract}

\section{INTRODUCTION}
Since the outbreak of COVID-19 at the end of 2019 due to a new virus (SARS-Cov-2) belonging to the family of coronavirus, the pandemic has significantly impacted daily life in the United States and rest of the world and continues to impact the global economy and human health. The first confirmed case of COVID-19 in the United States was identified on January 23, 2020 and the first death caused by COVID-19 in the United States is believed to have occurred on February 6, 2020 in Santa Clara County, California. The virus has grown and spread rapidly across the country since then. Solely in the United States, more than 44 million cases have been identified causing more than 700,000 deaths. As of October 1, 2021, COVID-19 has infected over $234$ million people and killed more than 4.5 million across the planet \cite{OWM}.

\subsection{Related Work}
Mathematical models for infectious diseases are considered an important tool for analyzing epidemiological features and transmission due to seminal contributions by Kermack \cite{kermack1927contribution} \cite{kermack1932contributions} \cite{kermack1933contributions}. A large number of studies on COVID-19 have been conducted by researchers and organizations around the world to estimate the dynamics of this infectious disease. Some of the most popular methods are SIR (Susceptible, Infected, and Recovered) and SEIR (Susceptible, Exposed, Infected, and Recovered). A time-window based SIR model was proposed \cite{liao2020tw}, for dynamic data analysis measuring the basic infection number and predicting the growth rate of the pandemic. A generalised SIR model \cite{singh2021generalized} was presented that incorporated multiple waves of daily reported cases and provided continuous predictions \& monitoring of the COVID-19 pandemic. Metapopulation Dynamics \cite{grenfell1997meta} \cite{hanski1998metapopulation} \cite{keeling2004metapopulation} was used to study and evaluate the effects of domestic and international travel limitations on the spread of COVID-19 \cite{chinazzi2020effect}. Mean-field theory \cite{fanelli2020analysis} was also applied to analyze and predict the spread of COVID-19 in countries such as China, Italy and France. A Bayesian Regression Model \cite{van2020aerosol} was used to analyze the aerosol stability of COVID-19. Recently, Deep Learning (DL) methods have also been at the forefront to analyze and predict the spread of COVID-19. \cite{chimmula2020time} was the first study that proposed using a Long Short-Term Memory (LSTM) network to model the spread of COVID-19 in Canada based on past data. One of the key features of this approach was considering the recovery rate during the development of the model. In A comparative study \cite{kirbacs2020comparative} was presented between Auto-Regressive Integrated Moving Average (ARIMA), Nonlinear Autoregression Neural Network (NARNN) and Long Short-Term Memory (LSTM) approaches to model and estimate the total number of COVID-19 infections for 8 different European countries. Four deep learning models: Long Short-term Memory (LSTM), Gated Recurrent Unit (GRU), Convolutional Neural Network (CNN) and Multivariate Convolutional Neural Network (MCNN) were compared and analyzed on the basis of consistency \& accuracy - with CNN outperforming other deep learning models \cite{nabi2021forecasting}.

\subsection{Objectives and Contributions}
The fundamental objective is to develop a data-driven physics-based approach which can be used for finite time estimation of  the spread of the COVID-19 pandemic across the United States. This finite estimation model can better estimate the growth of the pandemic for a shorter future time horizon compared to the existing methods. An early work on this topic was previously studied  in \cite{rastgoftar2021mass} where the authors assumed that evolution of the pandemic in every US state is not impacted by the growth in other US states. In contrast, this paper significantly advances this existing work by quantifying the inter-state influences and incorporating them into learning and prediction of the pandemic growth. To this end, our primary objective is to apply the temporal and spatial conservation laws to model and learn the evolution of the COVID-19 pandemic based on the total number of cases, deaths and recoveries across the United States. We then obtain the interstate influences and the parameters for the proposed dynamics by relying on the model, the pandemic data and leveraging quadratic programming \& neural networks.
The secondary objective of this article is to verify the stability of the pandemic. Using the open-source data from \cite{OWM}, we model the spread dynamics and use the stability criterion proposed in \cite{rastgoftar2021mass} to analyze the stability of the growth of the COVID-19 pandemic across the United States.

\subsection{Organization of the paper}
We detail the problem statement in Section II. In Section III, our proposed finite-time estimation framework is described. A novel approach to learning growth of the pandemic is provided in Section IV. We analyze the stability of the growth of the pandemic in Section V. Our evaluation and results are presented in Section VI. Finally, we provide discussion along with thoughts for future directions and conclude our work in Section VII.

\section{Problem Statement}
 We consider the growth of the COVID-19 pandemic in the United States where $50$ US states and District of Columbia are sorted alphabetically and identified by set $\mathcal{V}=\left\{1,\cdots,51\right\}$. We  define $t_i[k]$, $d_i[k]$, and $r_i[k]$ as the~total~number~of~cases, deaths, and recoveries, respectively, with $i\in \mathcal{V}$ at  day $k=1,\cdots,569$ from the establishment of pandemic growth. Estimations of $t_i[k]$, $d_i[k]$, and $r_i[k]$ are represented as $\hat{t}_i[k]$, $\hat{d}_i[k]$, and $\hat{r}_i[k]$, respectively. We apply \textit{temporal} and \textit{spatial} conservation laws to model and learn the evolution of the pandemic. The temporal conservation law is given by
 
\begin{equation}
    \label{MCLTemporalComponent}
    \hat{\mathbf{x}}_i[k+1]=\hat{\mathbf{x}}_i[k]+\hat{\mathbf{u}}_i[k],\qquad k=1,2,\cdots,569
\end{equation}

where $\hat{\mathbf{x}}_i[k]=\begin{bmatrix}\hat{t}_i[k]&\hat{d}_i[k]&\hat{r}_i[k]\end{bmatrix}^T$ is the state vector and $\hat{\mathbf{u}}_i[k]=\begin{bmatrix}\hat{u}_{t,i}[k]&\hat{u}_{d,i}[k]&\hat{u}_{r,i}[k]\end{bmatrix}^T$ is the input vector specifying number of new cases, new deaths, and new recoveries at day $k$. The spatial conservation law is given by

\begin{equation}
    \label{MCLSpatialComponent}
    \hat{a}_i[k]=\hat{t}_i[k]-\hat{d}_i[k]-\hat{r}_i[k],\qquad k=1,2,\cdots, 569.
\end{equation} 

where $\hat{a}_i[k]$ represents the number of active cases at day $k$.
This paper assumes that the number of new cases, new deaths, and new recoveries at day $k$ are proportional to the number of active cases over the past $N_\tau $ days from day  $k$. Therefore, we can define $\hat{u}_{t,i}$, $\hat{u}_{d,i}$, $\hat{u}_{r,i}$ as follows:

\begin{subequations}
\label{3333}

\begin{equation}
\label{2a}
\resizebox{0.99\hsize}{!}{%
$
\begin{split}
    \hat{u}_{t,i}[k]=&\beta[k]\sum_{h=1}^{N_\tau}\omega_{i,i,h}[k]\hat{a}_i[k-h]\\+
    &\left(1-\beta[k]\right)\sum_{h=1}^{N_\tau}\sum_{j\in \mathcal{V}}\omega_{i,j,h}[k]\hat{a}_j[k-h],\qquad \forall i\in \mathcal{V},
\end{split}
$
}
\end{equation}
\begin{equation}
\label{2b}
\resizebox{0.99\hsize}{!}{%
$
\begin{split}
     \hat{u}_{d,i}[k]=&\beta[k]\sum_{h=1}^{N_\tau}\lambda_{i,i,h}[k]\hat{a}_i[k-h]\\
   +&\left(1-\beta[k]\right)\sum_{h=1}^{N_\tau}\sum_{j\in \mathcal{V}}\lambda_{i,j,h}[k]\hat{a}_j[k-h],\qquad \forall i\in \mathcal{V},
\end{split}
   $
   }
\end{equation}
\begin{equation}
\label{2c}
\resizebox{0.99\hsize}{!}{%
$
\begin{split}
     \hat{u}_{r,i}[k]=&\beta[k]\sum_{h=1}^{N_\tau}\theta_{i,i,h}[k]\hat{a}_i[k-h]\\
     +&\left(1-\beta[k]\right)\sum_{h=1}^{N_\tau}\sum_{j\in \mathcal{V}}\theta_{i,j,h}[k]\hat{a}_j[k-h],\qquad \forall i\in \mathcal{V}.
\end{split}
$
}
\end{equation}
\end{subequations}

where $\omega_{i,j,h}[k]:\mathcal{V}\times \mathcal{V}\times\left\{1,\cdots,N_\tau\right\}\rightarrow \mathbb{R}_{\geq 0}$, $\lambda_{i,j,h}[k]:\mathcal{V}\times \mathcal{V}\times\left\{1,\cdots,N_\tau\right\}\rightarrow \mathbb{R}_{\geq 0}$, and $\theta_{i,j,h}[k]:\mathcal{V}\times \mathcal{V}\times\left\{1,\cdots,N_\tau\right\}\rightarrow \mathbb{R}_{\geq 0}$ assign the influence of the number of active cases in state $j\in \mathcal{V}$ in day $k-h$ on the number of new active cases, deaths, and recoveries in state $i\in \mathcal{V}$ at day $k$. Scaling parameter $\beta[k]\in \left[0,1\right]$. If $\beta[k]=1$, the second term on the right-hand side of Eqs. \eqref{2a}, \eqref{2b}, and \eqref{2c} vanishes and the pandemic independently grows in every US state. Otherwise ($\beta\in \left[0,1\right)$), evolution of pandemic in every US state is influenced by the growth in other US states. 

We first develop a finite time estimation model to predict the number of new cases, deaths, and recoveries in a finite number of future days and analyze the stability of growth of a pandemic disease in Section \ref{Finite-Time Estimation Model}. Then, we  offer an integration of quadratic programming and neural network to learn parameters $\omega_{i,j,h}[k]$, $\lambda_{i,j,h}[k]$, and $\theta_{i,j,h}[k]$ and $\beta[k]$ at every day $k$ and characterize the interstate influences on transmitting the current pandemic disease (see Section \ref{Learning of a Pandemic Growth}).

\section{Finite-Time Estimation of the Pandemic Growth}
\label{Finite-Time Estimation Model}
Per definition of the number of active cases, given in \eqref{MCLSpatialComponent}, the control input $\hat{\mathbf{u}}_i[k]$ is obtained as follows:
\begin{equation}
\label{uhattttttttttt}
    \hat{\mathbf{u}}_i[k]=\sum_{h=1}^{N_t}\sum_{j\in \mathcal{V}}\mathbf{G}_{i,j,h}[k]\hat{\mathbf{x}}[k-h],\qquad \forall i\in \mathcal{V},
\end{equation}
where 
\begin{equation}
    \mathbf{G}_{i,j,h}=\begin{bmatrix}\omega_{i,j,h}&-\lambda_{i,j,h}&-\theta_{i,j,h}\end{bmatrix} \otimes \mathbf{1}_{3\times 1}\in \mathbb{R}^{3\times 3}.
\end{equation}

We define 
\begin{subequations}
\begin{equation}
    \hat{\mathbf{y}}_i[k]=\begin{bmatrix}\hat{\mathbf{x}}_i^\mathsf{T}[k]&\cdots&\hat{\mathbf{x}}_i^\mathsf{T}[k-N_\tau+1]\end{bmatrix}^\mathsf{T}\in \mathbb{R}^{3N_\tau\times 1},
\end{equation}
\begin{equation}
   \hat{\mathbf{Y}}=\begin{bmatrix}\hat{\mathbf{y}}_1^\mathsf{T}&\cdots&\hat{\mathbf{y}}_{51}^\mathsf{T}\end{bmatrix}^\mathsf{T}\in \mathbb{R}^{153N_\tau \times 1},
\end{equation}

\begin{equation}
\label{lambdakkkkkkkkkk}
\resizebox{0.99\hsize}{!}{%
$
\mathbf{L}_{i,j}[k]=
\begin{bmatrix}
    \mathbf{0}&\mathbf{I}_3&\cdots&\mathbf{0}\\
    \vdots&\vdots&\ddots&\vdots\\
    \mathbf{0}&\mathbf{0}&\cdots&\mathbf{I}_3\\
    \mathbf{G}_{i,j,N_\tau}[k]&\mathbf{G}_{i,j,N_\tau-1}[k]&\cdots&\mathbf{I}+\mathbf{G}_{i,j,1}[k]\\
    \end{bmatrix}\in \mathbb{R}^{3N_\tau\times 3N_\tau}
    $
    }
\end{equation} 

\begin{equation}
    \mathbf{\Lambda}_0=
    \begin{bmatrix}
        \mathbf{L}_{1,1}[k]&\mathbf{L}_{1,2}[k]&\cdots&\mathbf{L}_{1,51}[k]\\
        \mathbf{L}_{2,1}[k]&\mathbf{L}_{2,2}[k]&\cdots&\mathbf{L}_{2,51}[k]\\
        \vdots&\vdots&\ddots&\vdots\\
        \mathbf{L}_{51,1}[k]&\mathbf{L}_{51,2}[k]&\ddots&\mathbf{L}_{51,51}[k]\\
    \end{bmatrix}
\end{equation}
\begin{equation}
    \mathbf{\Lambda}_1=\begin{bmatrix}
        \mathbf{L}_{1,1}[k]&0&\cdots&0\\
        0&\mathbf{L}_{2,2}[k]&\cdots&0\\
        \vdots&\vdots&\ddots&\vdots\\
        0&0&\ddots&\mathbf{L}_{51,51}[k]\\
    \end{bmatrix}
\end{equation}
\begin{equation}\label{Lmatrix}
    \mathbf{L}[k]=\beta[k]\mathbf{\Lambda}_1
    +\left(1-\beta[k]\right)\mathbf{\Lambda}_0
\end{equation}
\end{subequations}
where ``$\otimes$'' is the Kronecker product symbol, $\mathbf{I}_3\in \mathbb{R}^{3\times 3}$ is the identity matrix, $\mathbf{1}_{3\times 1}\in \mathbb{R}^{3\times 1}$ is the one-entry vector. Then, the estimation of the pandemic growth is obtained by the following  $M$-step prediction dynamics:
\begin{equation}\label{daykmodel}
\resizebox{0.99\hsize}{!}{%
$
\begin{cases}
    \hat{\mathbf{Y}}[k+M]=
    \mathbf{L}^M[k]\hat{\mathbf{Y}}[k]\\
     \hat{\mathbf{x}}_i\left[k+M\right]=
    \begin{bmatrix}
        \mathbf{0}_{M_1\times 3}&\mathbf{I}_3&\mathbf{0}_{M_2\times 3}
    \end{bmatrix}\hat{\mathbf{Y}}[k+M]
\end{cases}
,\qquad i\in \mathcal{V}
$
}
\end{equation}
where
\begin{subequations}
\begin{equation}
    M_1=\left((i-1)N_\tau+3(M-1)\right)
\end{equation}
\begin{equation}
    M_2=\left(51-i\right)N_\tau+3\left(N_\tau-M\right)
\end{equation}
\end{subequations}
and  $\hat{\mathbf{x}}_i\left[k+M\right]=\begin{bmatrix}
        \hat{t}_i[k+M]&\hat{d}_i[k+M]&\hat{r}_i[k+M]
    \end{bmatrix}
    ^\mathsf{T}$ aggregates the prediction of day $k$ for the the total number of cases, deaths, and recoveries at day $k+M$. 

\section{Learning of the Pandemic Growth}
\label{Learning of a Pandemic Growth}
The parameters $\omega_{i,j,h}[k]:\mathcal{V}\times \mathcal{V}\times\left\{1,\cdots,N_\tau\right\}\rightarrow \mathbb{R}_{\geq 0}$, $\lambda_{i,j,h}[k]:\mathcal{V}\times \mathcal{V}\times\left\{1,\cdots,N_\tau\right\}\rightarrow \mathbb{R}_{\geq 0}$, and $\theta_{i,j,h}[k]:\mathcal{V}\times \mathcal{V}\times\left\{1,\cdots,N_\tau\right\}\rightarrow \mathbb{R}_{\geq 0}$ must be all non-negative at every day $k$ since the number of new cases, deaths, and recoveries are positively impacted by the number of active cases. Therefore, matrices $\mathbf{\Lambda}_0$ and  $\mathbf{\Lambda}_1$ are non-negative. Matrices $\mathbf{\Lambda}_0$ and $\mathbf{\Lambda}_1$ are learned by solving quadratic programming problem defined in Sections \ref{MatrixL0} and \ref{MatrixL1}, respectively. To characterize the interstate influences, parameter $\beta$ is learned by a neural network presented in Section \ref{beta}.

\subsection{Matrix $\mathbf{\Lambda}_0$}
\label{MatrixL0}
In this section, we assume that the pandemic grows under quarantined condition in every US state $i\in \mathcal{V}$. Therefore, the spread of pandemic in a particular state is independent of other US states. Therefore, $\beta[k]=1$; the second term on the right-hand side of Eq. \eqref{Lmatrix} vanishes; and matrix $\mathbf{L}=\mathbf{\Lambda}_1$ is obtained by minimizing cost function
\begin{equation}
\resizebox{0.99\hsize}{!}{%
$
    C_q[k]={\frac{1}{2}}\left({\mathbf{\Lambda}}_1[k]\hat{\mathbf{Y}}[k]-\mathbf{Y}[k]\right)^\mathsf{T}\mathbf{W}_q\left({\mathbf{\Lambda}}_1\hat{\mathbf{Y}}[k]-\mathbf{Y}[k]\right)
$
}
\end{equation}
subject to the $\omega_{i,i,h}\geq0$, $\lambda_{i,i,h}\geq0$, and $\theta_{i,i,h}\geq 0$ for every $i\in \mathcal{V}$ and $h\in \left\{1,\cdots,N_\tau\right\}$, where $\mathbf{Y}[k]$ aggregates the actual number of infected cases, deaths, and recoveries recorded at days $k-N_\tau+1$ through $k$ and $\mathbf{W}_q\in \mathbb{R}^{153N_\tau\times 153N_\tau}$ is a positive definite weight matrix. Note that cost function $C_q$ is quadratic with respect to $\omega_{i,i,h}$, $\lambda_{i,i,h}$, and $\theta_{i,i,h}$. Therefore, matrix $\mathbf{L}=\mathbf{\Lambda}_1$ is assigned by solving a quadratic programming problem.

\subsection{Matrix $\mathbf{\Lambda}_1$}
\label{MatrixL1}
In this section,  we assume that pandemic grows under non-quarantined conditions in every US state $i\in \mathcal{V}$. Therefore, the growth of pandemic disease in every US state $i\in \mathcal{V}$ is influenced by other US states and $\beta[k]=0$, the first term on the right-hand side of Eq. \eqref{Lmatrix} vanishes, and $\mathbf{L}[k]=\mathbf{\Lambda}_0[k]$.Therefore, matrix $\mathbf{L}=\mathbf{\Lambda}_0$ is determined by minimizing quadratic cost function
\[
\resizebox{0.99\hsize}{!}{
$
C_{n}[k]={\frac{1}{2}}\left({\mathbf{\Lambda}}_0[k]\hat{\mathbf{Y}}[k]-\mathbf{Y}[k]\right)^\mathsf{T}\mathbf{W}_n\left({\mathbf{\Lambda}}_0\hat{\mathbf{Y}}[k]-\mathbf{Y}[k]\right)
$
}
\]
subject to the $\omega_{i,j,h}\geq0$, $\lambda_{i,j,h}\geq0$, and $\theta_{i,j,h}[k]\geq 0$ for every $i,j\in \mathcal{V}$ and $h\in \left\{1,\cdots,14\right\}$, where $\mathbf{W}_n\in \mathbb{R}^{153N_\tau\times 153N_\tau}$ is a positive definite weight matrix. Similar to the quarantined condition, matrix $\mathbf{L}=\mathbf{\Lambda}_1$  can be assigned by solving a quadratic programming problem.

\subsection{Parameter $\beta$}
\label{beta}

Vector $\mathbf{Y}[k]$ is the input to the network, which consists of the total number of cases, deaths, and recoveries of all US states and District of Columbia for the past  $N_\tau$ days from day $k$. The fully connected hidden layer consists of $51$ neurons. Defining $\beta \in (0,1)$ as the output of the neural network, we use the existing statistics  of COVID-19 in the US states, recorded from March 12, 2020 to October 1, 2021 to train and test the neural network. In the case of the state of Vermont,  $\beta=1$, from March 12,  2020 until October 1, 2020 because of the spread of COVID was controlled to an extent. From October 2, 2020 until April 23, 2021, the state of Vermont had an increase in cases and deaths over this period. Therefore, we have $\beta=0$ from October 2, 2020 till April 23, 2021 and the statistics recorded over this period are used to train the network. The data from April 24, 2021 until October 1, 2021 was used to test the network and obtain the estimated number of cases, deaths and recoveries.

\section{Stability of the pandemic growth}

Following the stability analysis first provided in \cite{rastgoftar2021mass}, we define active cases, originally given by \eqref{MCLSpatialComponent} as follows:
\begin{equation}
    {a}_i[k]=\mathbf{a}\hat{\mathbf{x}}_i[k],\qquad k=1,2,\cdots,569 
\end{equation}
where 
\begin{equation}
\label{adefinition1}
    \mathbf{a}=\begin{bmatrix}1&-1&-1\end{bmatrix}\in \mathbb{R}^{1\times 3}
\end{equation}
From \eqref{uhattttttttttt}, we can see that 
\begin{equation}
\label{gk}
    \mathbf{G}_{i,j,h}[k]=\mathbf{K}_{i,j,h}[k]\mathbf{a}
\end{equation}where 
\begin{equation}
\label{adefinition}
    \mathbf{K}_{i,j,h}[k]=\begin{bmatrix}\omega_{i,j,h}[k]&\lambda_{i,j,h}[k]&\theta_{i,j,h}[k] \end{bmatrix}^T
\end{equation}

Using \eqref{gk}, we can rewrite \eqref{uhattttttttttt} as
\begin{equation}
\label{newuhat}
    \hat{\mathbf{u}}_i[k]=\sum_{h=1}^{N_t}\sum_{j\in \mathcal{V}}\mathbf{K}_{i,j,h}[k]\hat{\mathbf{a}}[k-h],\qquad \forall i\in \mathcal{V},
\end{equation}

Substituting \eqref{newuhat} in \eqref{MCLTemporalComponent} and premultiplying both sides by $\mathbf{a}$, we get 

\begin{equation}
\resizebox{0.99\hsize}{!}{$
\label{ak}
    \mathbf{a}\hat{\mathbf{x}}[k]=\mathbf{a}\hat{\mathbf{x}}[k-1] + \sum_{h=1}^{N_t}\sum_{j\in \mathcal{V}}\mathbf{a}\mathbf{K}_{i,j,h}[k]\hat{\mathbf{a}}[k-h]\qquad \forall i\in \mathcal{V}
    $}
\end{equation}

Let
\begin{equation}
    \gamma_{i,j,h}[k]:=\mathbf{a}\mathbf{K}_{i,j,h}[k]=\omega_{i,j,h}[k]-\lambda_{i,j,h}[k]-\theta_{i,j,h}[k]
\end{equation}

Hence the terms in \eqref{ak} can be replaced to obtain the dynamics of the spread of active cases given by the following equation 

\begin{equation}
\label{differenceeqn}
    a[k]=\left(1+\gamma_{i,j,1}[k]\right)a[k-1]+\cdots+\gamma_{i,j,N_\tau}[k]a[k-N_\tau]
\end{equation}

Using the dynamics of the spread of active cases given by \eqref{differenceeqn}, the growth of the total number of active cases,  defined by \eqref{MCLSpatialComponent}, will reach stability if there exists a day $k_s$ such that eigenvalues of 

\begin{equation}
\resizebox{0.99\hsize}{!}{$
    \mathbf{\Gamma}[k]=
\begin{bmatrix}
    {0}&1&\cdots&{0}\\
    \vdots&\vdots&\ddots&\vdots\\
    {0}&{0}&\cdots&1\\
    \gamma_{i,j,N_\tau}[k]&\gamma_{i,j,N_\tau-1}[k]&\cdots&1+\gamma_{i,j,1}[k]\\
    \end{bmatrix} \\
    \in \mathbb{R}^{N_\tau\times N_\tau}
    $}
\end{equation}
are all inside the unit disk centered at the origin at every day $k>k_s$.

\section{Results}

\begin{figure}[h]
    \centering
    \subfigure[]{\includegraphics[width=0.49\linewidth]{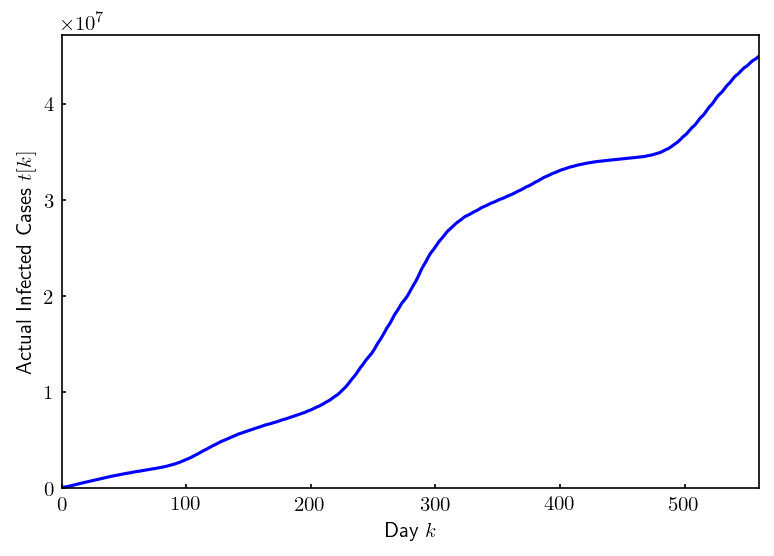}}
    \subfigure[]{\includegraphics[width=0.49\linewidth]{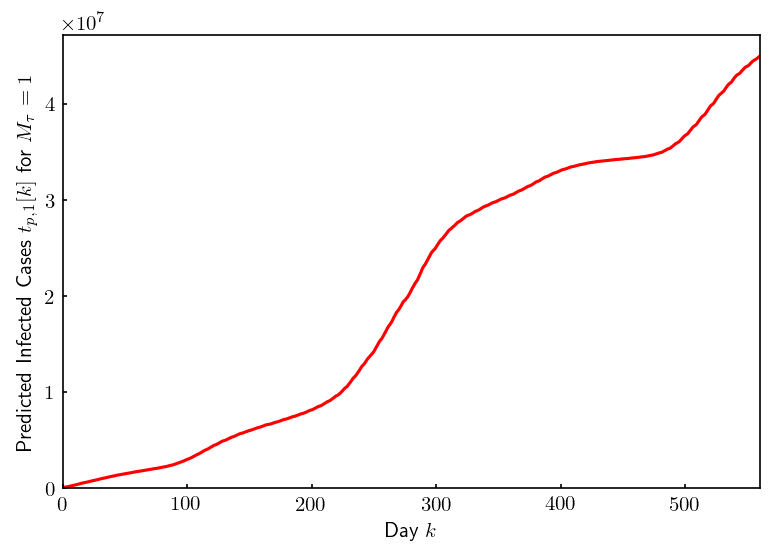}}
    \subfigure[]{\includegraphics[width=0.49\linewidth]{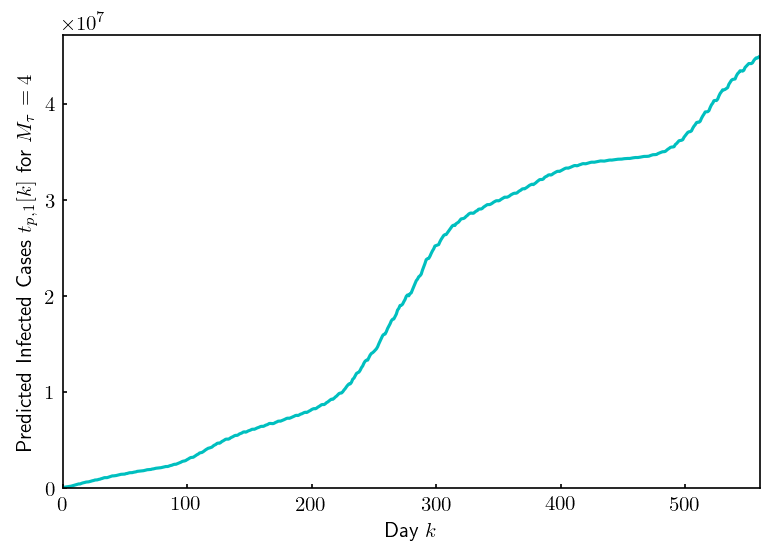}}
    \subfigure[]{\includegraphics[width=0.49\linewidth]{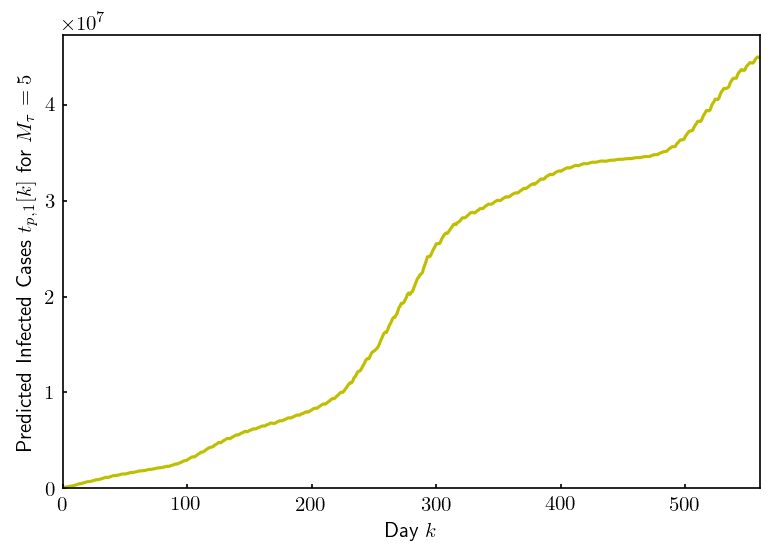}}
    \caption{For $k = 15,\cdots,569$ (a) Actual number of cases. Predicted number of cases for (b) $M_{\tau} = 1$, (c) $M_{\tau} = 4$, (d) $M_{\tau} = 5$.}
    \label{fig:allinfectedcases}
\end{figure}

Before proceeding to evaluate the effectiveness of our proposed finite-time estimation approach, we define the following relative error formulations:

\begin{equation}
    e_{t, M_{\tau}}[k] = \frac{t_{p, M_{\tau}}[k] - t[k]}{t[k]}
\end{equation}

\begin{equation}
    e_{d, M_{\tau}}[k] = \frac{d_{p, M_{\tau}}[k] - d[k]}{d[k]}
\end{equation}

\begin{equation}
    e_{r, M_{\tau}}[k] = \frac{r_{p, M_{\tau}}[k] - r[k]}{r[k]}
\end{equation}

The data is collected from \cite{OWM}. The spread of COVID-19 is considered over a period of 569 days from March 12, 2020 to October 1, 2021. For $k = 1,\cdots,569$, the total number of cases, deaths and recoveries are represented as $t[k]$, $d[k], r[k]$. We chose $N_{\tau} = 14$ for all of our experiments, to obtain the parameters of the model described in Section IV. Therefore, $t[k]$, $d[k], r[k]$ for $k = 1,\cdots,14$ (March 12, 2020 to March 25, 2020) are used to learn the parameters of the proposed model for $k \geq 15$ by solving the optimization problem defined in Section IV subject to constraints mentioned. We obtain gains $K_{1}[k],\cdots,K_{N_{\tau}}[k]$ for day $k$. These gains are then used in predicting the growth in total number of cases, deaths and recoveries from March 26, 2020. 

\subsection{Estimation and Prediction in the United States}

\begin{figure}[h]
    \centering
    \subfigure[]{\includegraphics[width=0.49\linewidth]{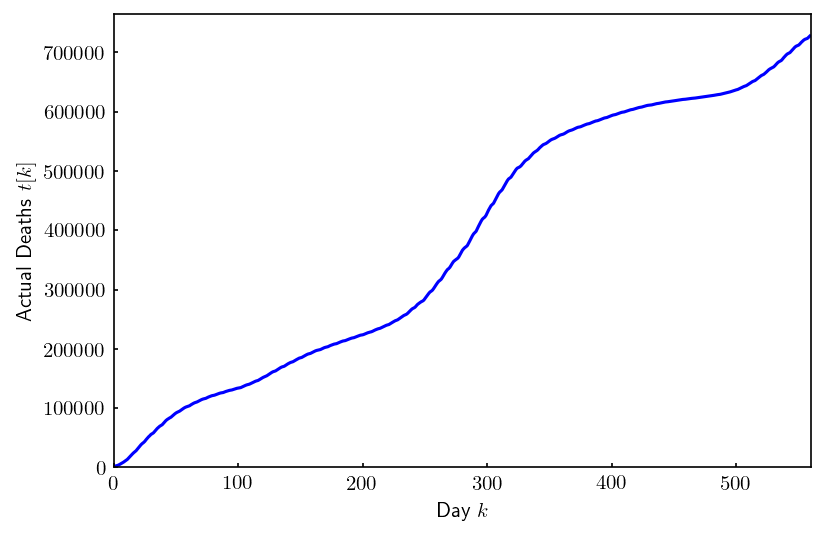}}
    \subfigure[]{\includegraphics[width=0.49\linewidth]{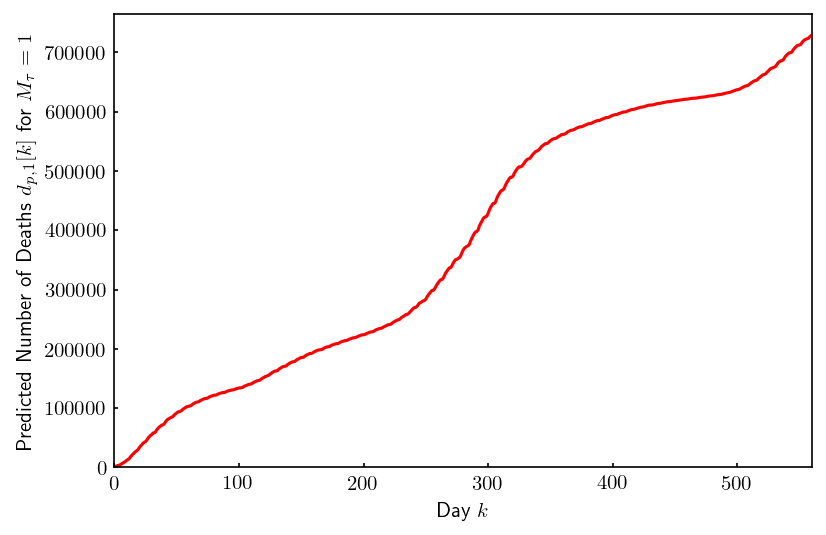}}
    \subfigure[]{\includegraphics[width=0.49\linewidth]{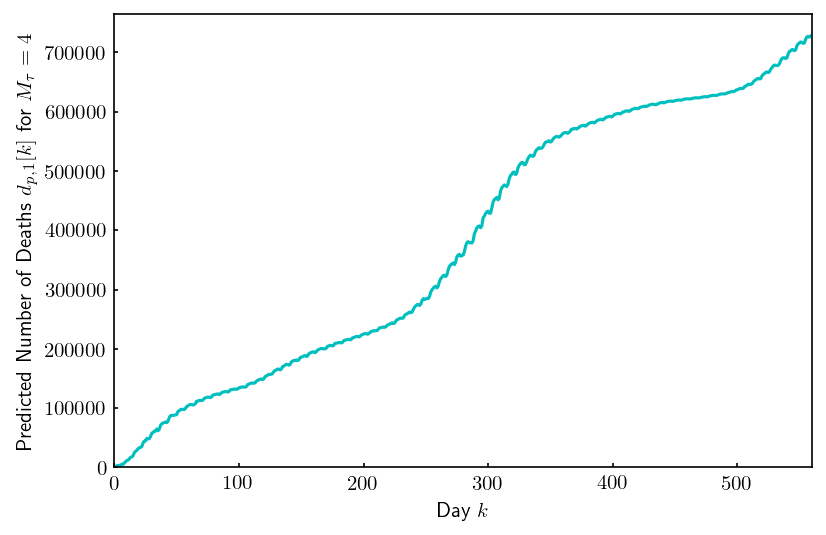}}
    \subfigure[]{\includegraphics[width=0.49\linewidth]{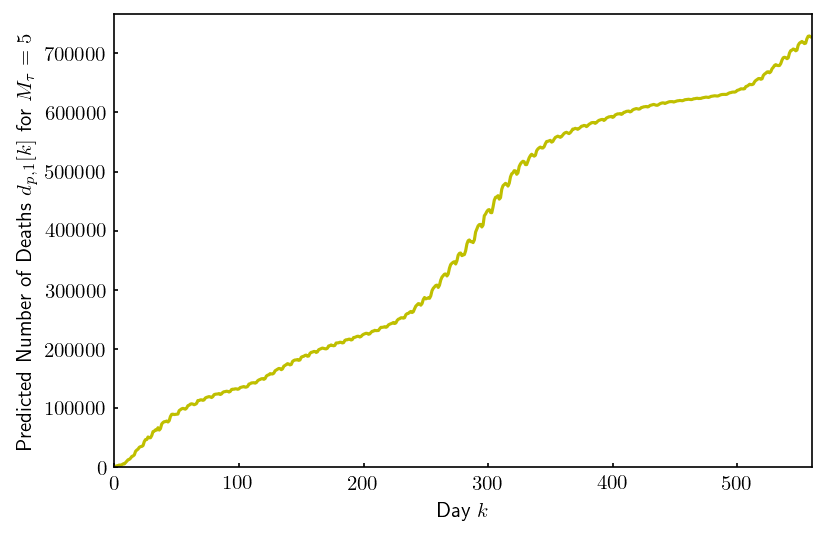}}
    \caption{For $k = 15,\cdots,569$ (a) Actual number of deaths. Predicted number of deaths for (b) $M_{\tau} = 1$, (c) $M_{\tau} = 4$, (d) $M_{\tau} = 5$.}
    \label{fig:alldeaths}
\end{figure}

\begin{figure}[h]
    \centering
    \subfigure[]{\includegraphics[width=0.49\linewidth]{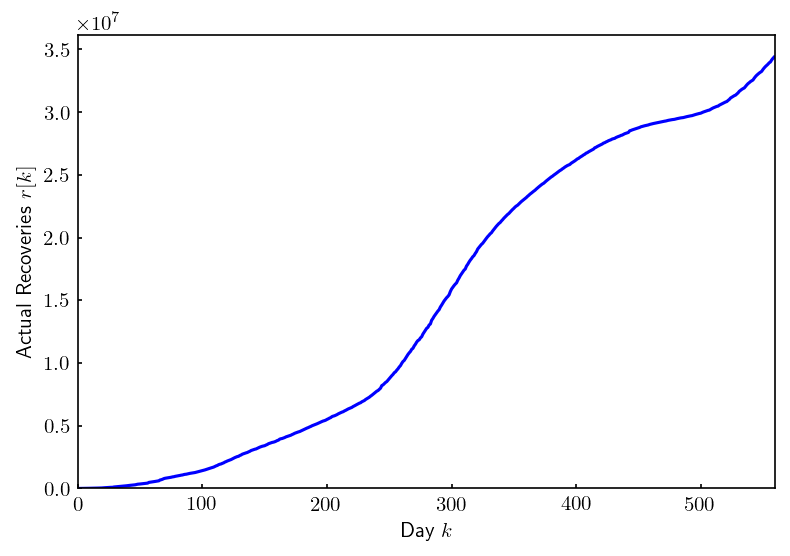}}
    \subfigure[]{\includegraphics[width=0.49\linewidth]{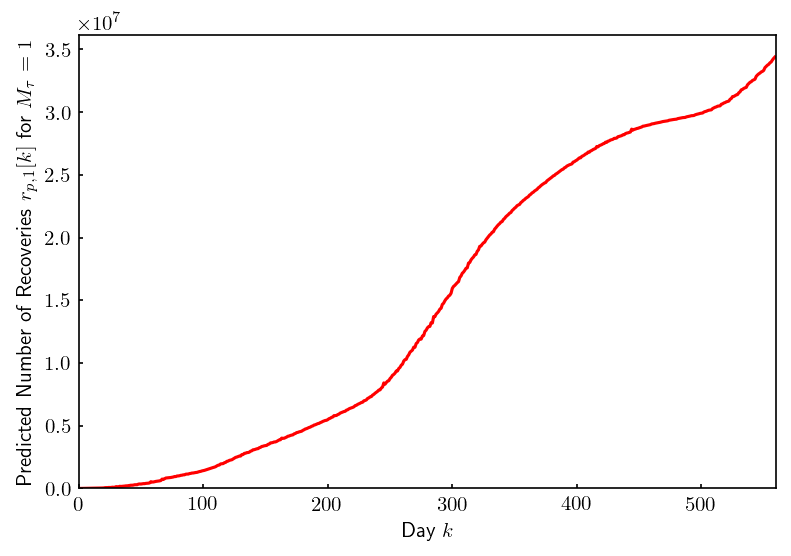}}
    \subfigure[]{\includegraphics[width=0.49\linewidth]{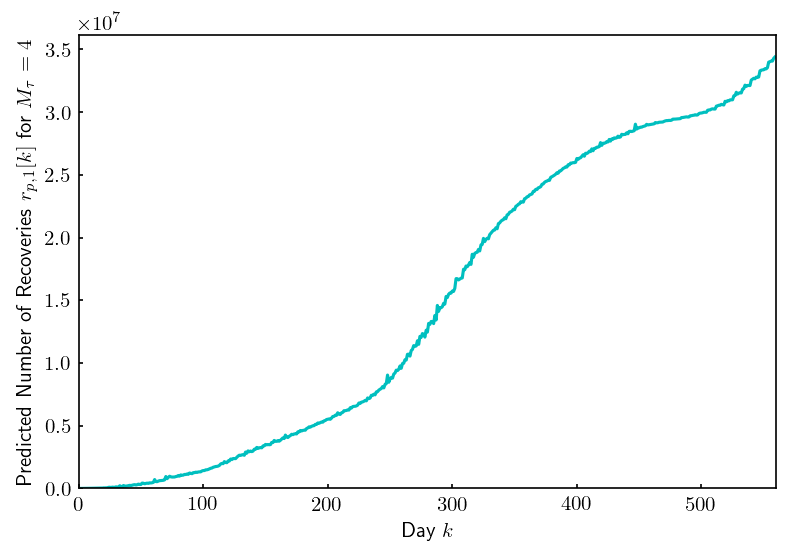}}
    \subfigure[]{\includegraphics[width=0.49\linewidth]{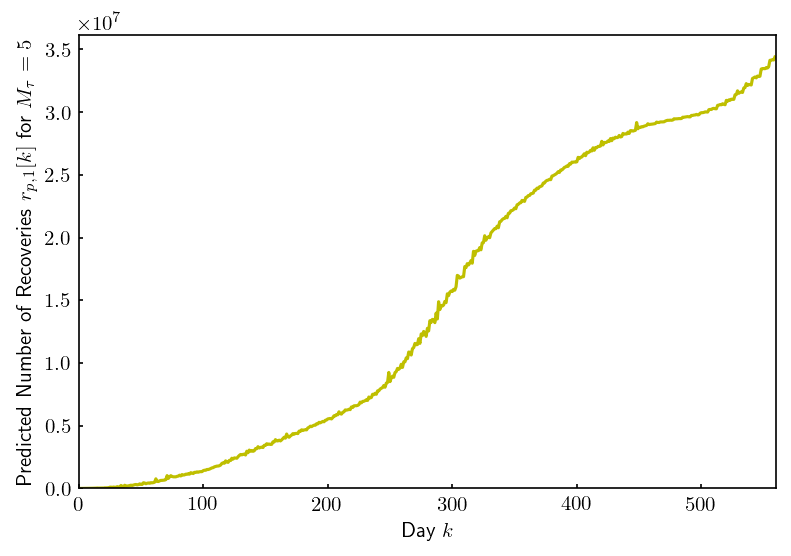}}
    \caption{For $k = 15,\cdots,569$ (a) Actual number of recoveries. Predicted number of recoveries for (b) $M_{\tau} = 1$, (c) $M_{\tau} = 4$, (d) $M_{\tau} = 5$.}
    \label{fig:allrecoveries}
\end{figure}

\begin{figure}[h]
    \centering
    \subfigure[]{\includegraphics[width=0.8\linewidth]{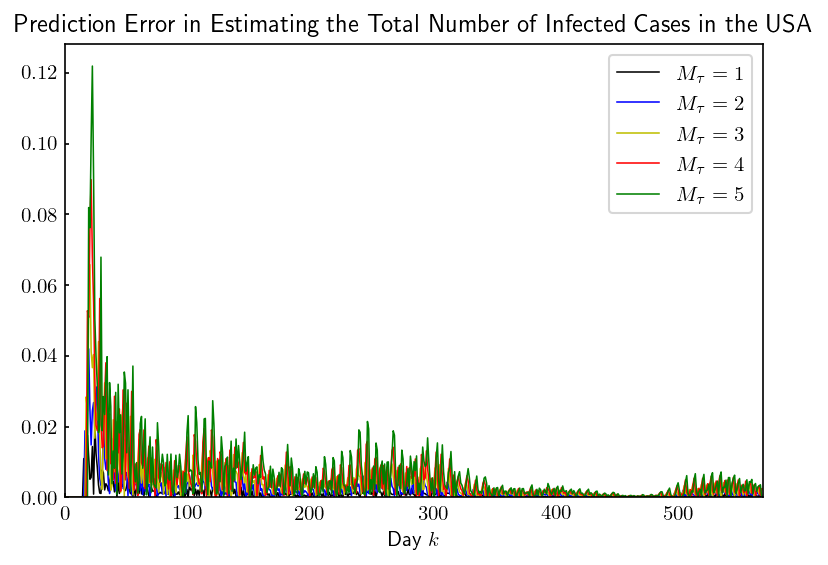}}
    \subfigure[]{\includegraphics[width=0.8\linewidth]{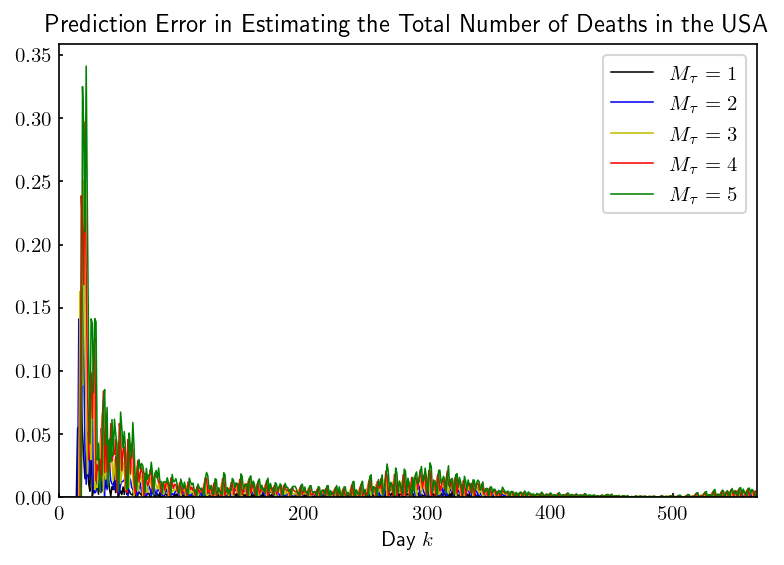}}
    \subfigure[]{\includegraphics[width=0.8\linewidth]{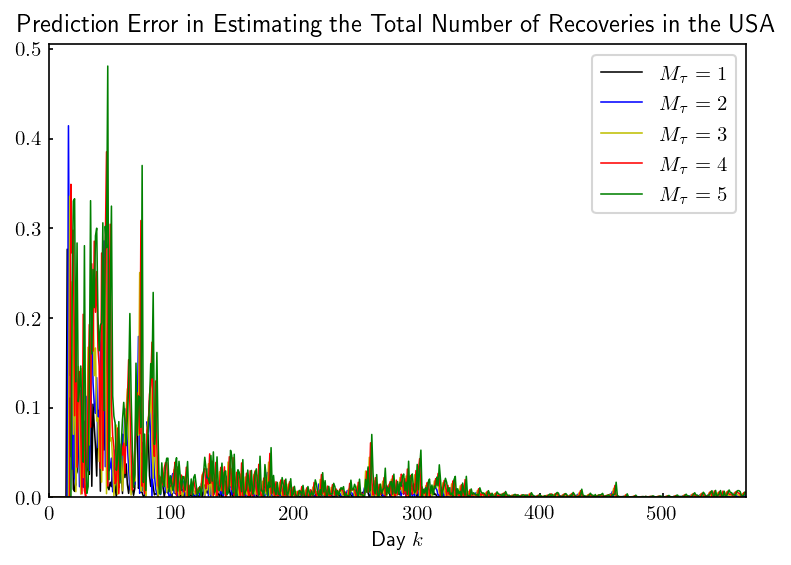}}
    \caption{Relative Prediction Errors in estimating the total number of (a) cases, (b) deaths, and (c) recoveries in US.}
    \label{fig:predictionerrors}
\end{figure}

\begin{figure}[h]
    \centering
    \includegraphics[width=0.90\linewidth]{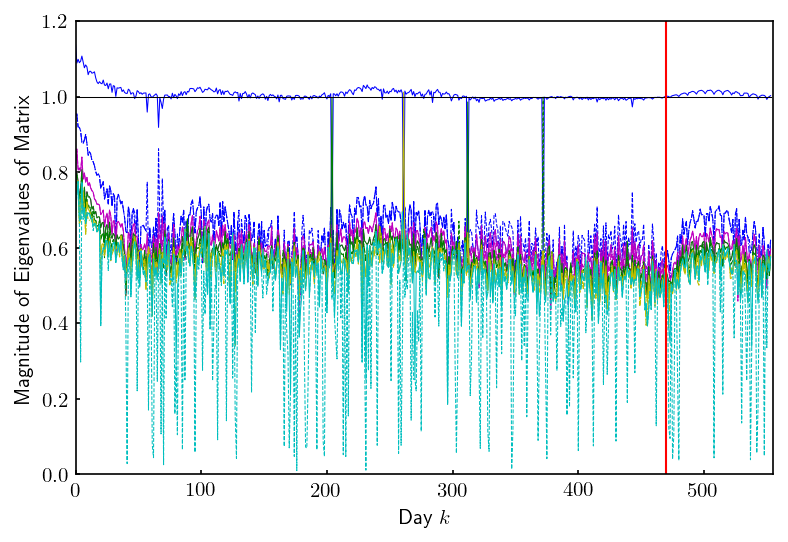}
    \caption{Magnitude of Eigenvalues of Matrix}
    \label{fig:stabilitymatrix}
\end{figure}

Figure \ref{fig:allinfectedcases}, \ref{fig:alldeaths}, \ref{fig:allrecoveries} shows the plots of actual and predicted number of cases, deaths and recoveries in the United States for $k=15,\cdots,569$. Figure \ref{fig:predictionerrors} shows the relative errors in total number of cases, deaths and recoveries in the US for $M_{\tau}=1,\cdots,5$ from which we can verify that the errors are less than $1.0\%$ 

Figure \ref{fig:stabilitymatrix} shows the plot between eigenvalues of matrix $\mathbf{\Gamma}[k]$ and $k$. The first eigenvalue of matrix $\mathbf{\Gamma}[k]$ is considered to be the eigenvalue which has the largest magnitude. We can see from the plot that the magnitude of the first eigenvalue fluctuates around 1 while the magnitudes of the remaining eigenvalues are all between 0 and 1 for $k=15,\cdots,469$. Around $k=220$, we first observe that the magnitude of the first eigenvalue is greater than 1. This is consistent with the status of the pandemic growth in the United States while approaching the end of october. From $k=470$, we again observe the same behaviour which approximately coincides with the spread of delta variant in the United States. Hence, we can conclude that the growth of total number of cases, deaths and recoveries has not reached stability.

\subsection{Estimation and Prediction in the state of New York}

\begin{figure}[h]
    \centering
    \subfigure[]{\includegraphics[width=0.49\linewidth]{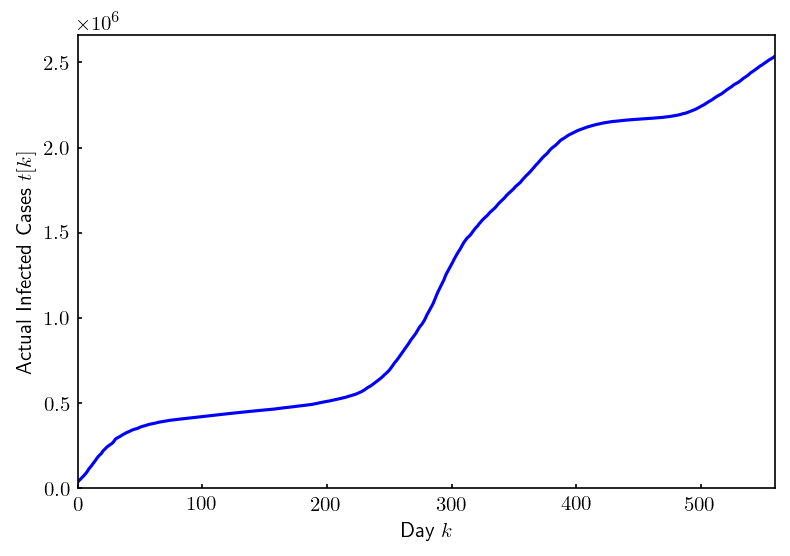}}
    \subfigure[]{\includegraphics[width=0.49\linewidth]{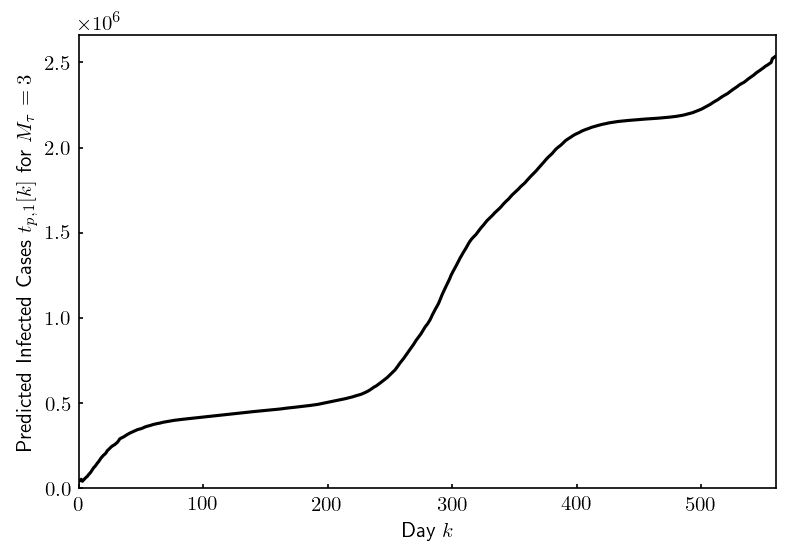}}
    \subfigure[]{\includegraphics[width=0.49\linewidth]{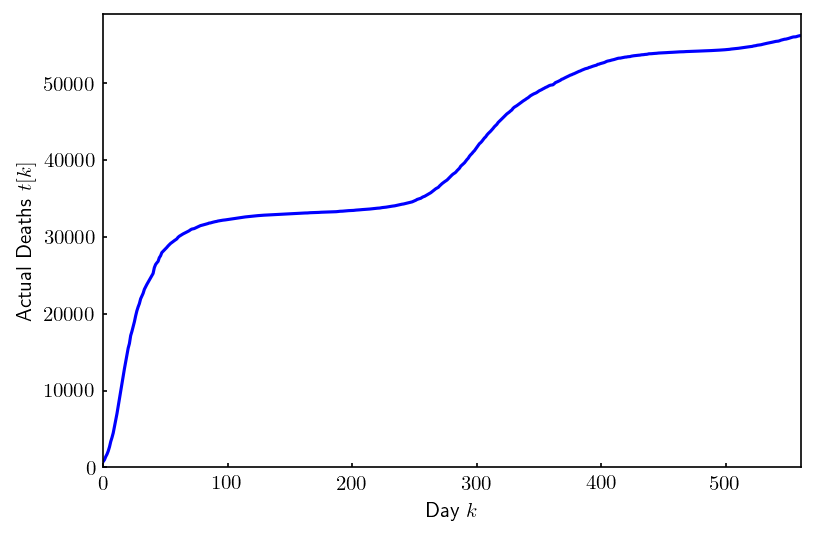}}
    \subfigure[]{\includegraphics[width=0.49\linewidth]{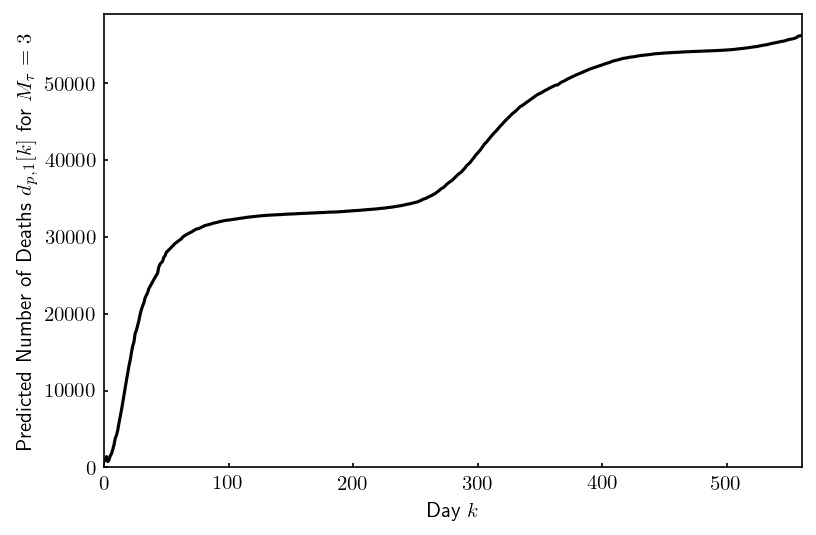}}
    \subfigure[]{\includegraphics[width=0.49\linewidth]{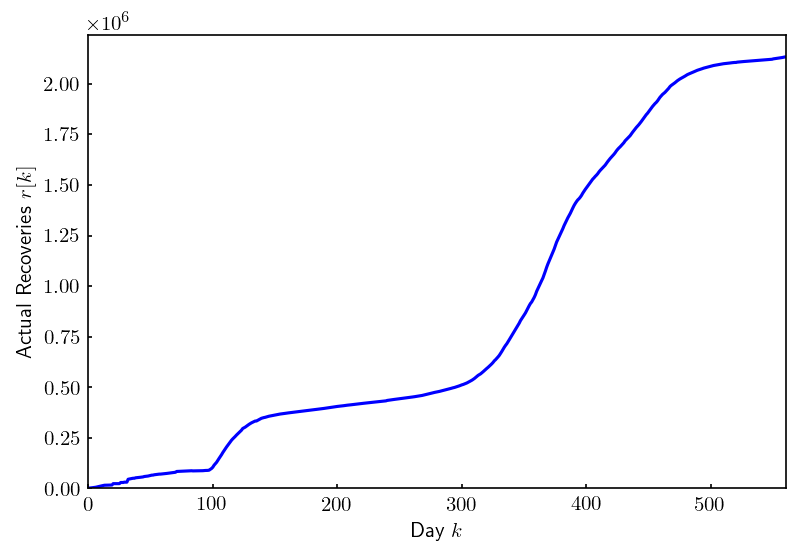}}
    \subfigure[]{\includegraphics[width=0.49\linewidth]{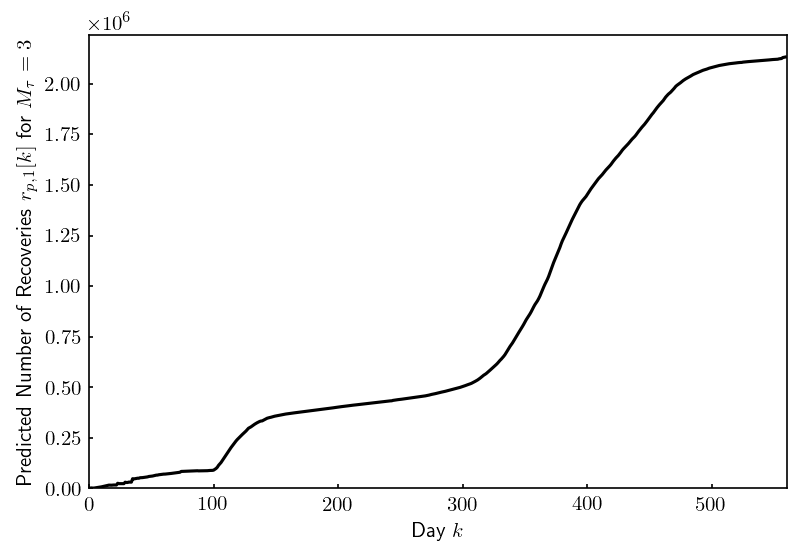}}
    \caption{For the state of New York, (a) actual number of cases reported, (b) predicted number of cases for $M_{\tau} = 3$, (c) total deaths reported, (d) predicted number of deaths for $M_{\tau} = 3$, (e) total recoveries reported, (f) predicted number of recoveries for $M_{\tau} = 3$.}
    \label{fig:nycombined}
\end{figure}

\begin{figure}[h]
    \centering
    \subfigure[]{\includegraphics[width=0.85\linewidth]{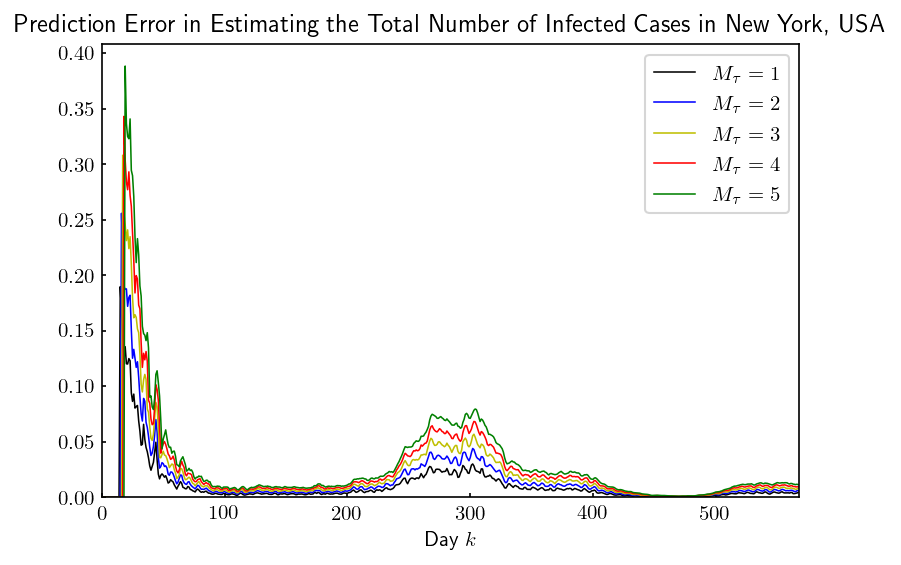}}
    \subfigure[]{\includegraphics[width=0.85\linewidth]{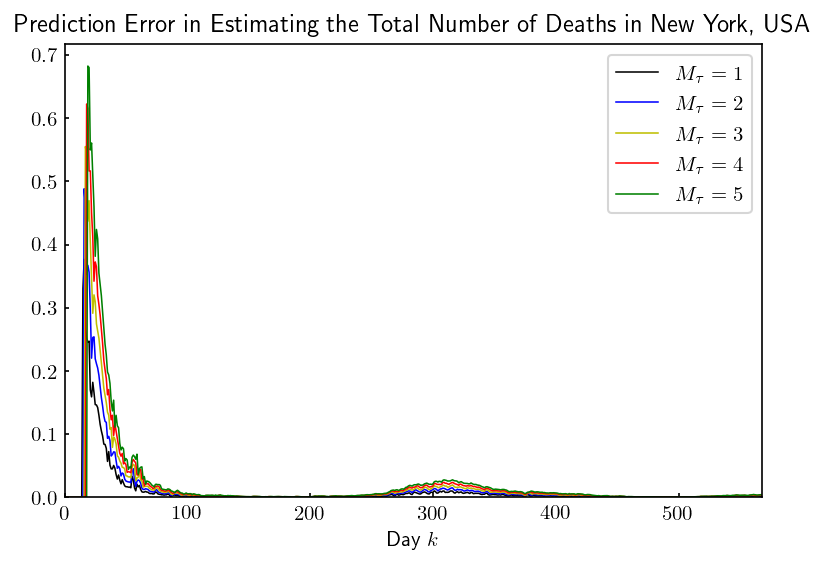}}
    \subfigure[]{\includegraphics[width=0.85\linewidth]{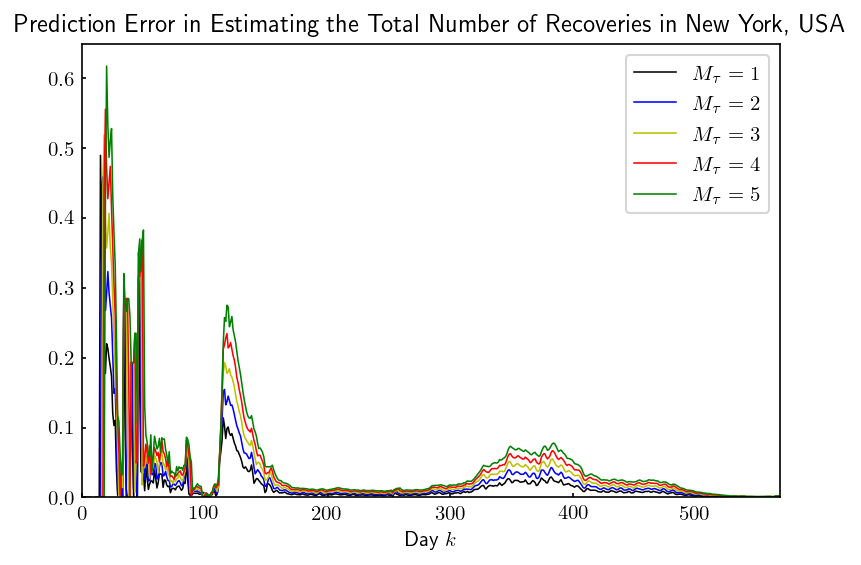}}
    \caption{Relative Errors in estimating the total number of (a) cases, (b) deaths, and (c) recoveries in New York. }
    \label{fig:predictionerrorsny}
\end{figure}

Let $\beta = 0$ i.e. the total number of cases, deaths and recoveries in the state of New York is dependent on total number of cases, deaths and recoveries in other 49 states, District of Columbia and New York itself. We test the approach described in Section III particularly for the state of New York because it is one of the most affected states in US due to the pandemic. Therefore, for $k=15,\cdots,569$, we use $t[k], d[k], r[k]$ from all 50 states and District of Columbia to predict and estimate the total number of cases, deaths and recoveries in New York. Figure \ref{fig:nycombined} shows the plots of actual and predicted number of infected cases, deaths and recoveries in New York for $k=15,\cdots,569$ for $M_{\tau}=3$. Figure \ref{fig:predictionerrorsny} plots the relative errors in predicting the number of cases,deaths and recoveries in the state of New York at day $k = 15,...,569$ for $M_{\tau}=3$. It can be seen that the relative errors in our estimations are less than $1\%$.

\subsection{Estimation and Prediction of Non-Quarantined or Quarantined state using Neural Network}

\begin{figure}[h]
    \centering
    \includegraphics[width=0.90\linewidth]{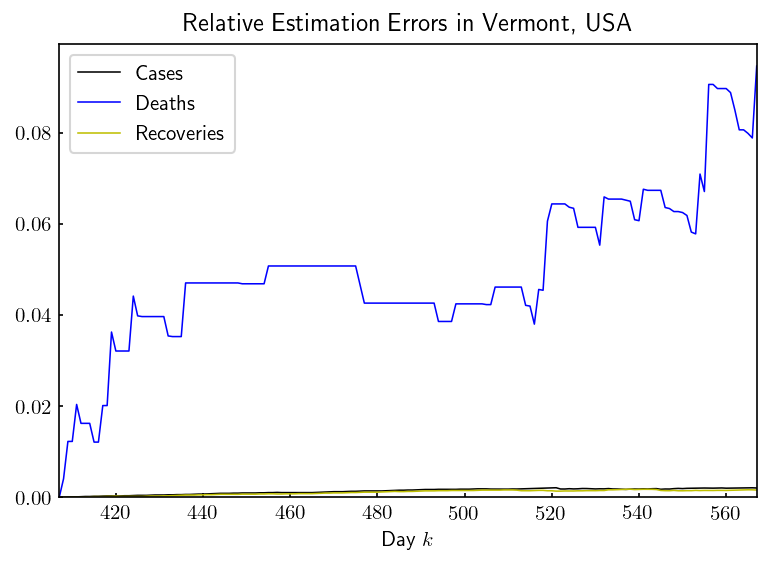}
     \caption{Relative Errors in estimating the total number of cases, deaths, and recoveries in Vermont.}
    \label{fig:predictionerrorsvermont}
\end{figure}

The objective of this experiment is to determine whether the US state of Vermont is in a quarantined ($\mathbf{\beta} = 1$) or non-quarantined ($\mathbf{\beta} = 0$) state using Neural Networks. As described in Section IV, the input to the Neural Network is $t[k], d[k], r[k]$ for the past $k= i-14,\cdots,i$ of all 50 states making the input of shape 3x14x51 for a particular day $k$. The hidden layer is a fully connected layer with ReLU activation function. The output from the Neural Network is the value $\beta$.  Using \eqref{2a} \eqref{2b} \eqref{2c}, we obtain the estimated number of new cases, deaths, and recoveries. Finally using \eqref{MCLTemporalComponent} we get the predicted number of cases, deaths and recoveries. Using this approach, we can see that the relative estimation errors are less than $0.01\%$ as shown in Figure \ref{fig:predictionerrorsvermont}. 

\section{Conclusion and Future Work}

We presented a new physics-based data-driven approach to model and predict the spread of COVID-19 pandemic in the United States. Using a time-sliding window and historical data, we described an algorithm to learn the parameters of the proposed approach and developed a prediction model using spatial and temporal conservation laws, to estimate in finite-time, in a receding-time horizon. The accuracy of the prediction model is established by estimating the number of cases, deaths and recoveries in United States and also in the state of New York from March 12, 2020 to October 1, 2021 with our relative estimation errors being less than $1\%$ during this time period. 

We also analyzed the stability of the growth of COVID-19 pandemic. Our results are coinciding with 
\begin{enumerate}
    \item the growth of the pandemic considering the devastating effects and spread of the delta variant since May 2021,
    \item the actual data reported (Figure \ref{fig:allinfectedcases}(a)) in which we can see the another wave i.e. an increase in the number of cases.   
\end{enumerate}

In future, we plan on using Markov Decision Process (MDP) to model the spread of the COVID-19 pandemic under the finite-time estimation approach described in this article. Assuming full-state observability, this novel decision-making model could potentially be used to effectively make state-wide and nation-wide recommendations by optimizing for non-pharmaceutical actions. 

\section{Acknowledgement}
This work has been supported by the National Science
Foundation under Award Nos. 2133690 and 1914581. 

\bibliographystyle{IEEEtran}
\bibliography{myref}

\end{document}